\documentclass[sigconf,natbib=true]{acmart} %

\usepackage{lipsum}

\AtBeginDocument{%
  \providecommand\BibTeX{{%
    \normalfont B\kern-0.5em{\scshape i\kern-0.25em b}\kern-0.8em\TeX}}}

\copyrightyear{2023}
\acmYear{2023}
\acmConference[Gen-IR@SIGIR2023]{The First Workshop on Generative Information Retrieval}{July 27, 2023}{Taipei, Taiwan}
\acmBooktitle{The First Workshop on Generative Information Retrieval (Gen-IR@SIGIR23), July 27, 2023, Taipei, Taiwan}
\acmDOI{}
\acmPrice{}
\acmISBN{}
\setcopyright{rightsretained}

\usepackage[utf8]{inputenc}
\usepackage[show]{chato-notes}
\usepackage{booktabs} %
\usepackage{amsmath}
\usepackage{multirow}
\usepackage{placeins}
\usepackage{wrapfig}
\usepackage{subfig}
\usepackage{makecell}
\usepackage{algorithm}
\usepackage{enumitem}
\usepackage{algpseudocode}

\usepackage{pgfplots}
\DeclareUnicodeCharacter{2212}{−}
\usepgfplotslibrary{groupplots,dateplot}
\usetikzlibrary{patterns,shapes.arrows}
\pgfplotsset{compat=newest}

\graphicspath{ {./graphics/} }
\usepackage{amsmath}

\newcommand{\craig}[1]{\textcolor{black}{#1}}
\newcommand{\se}[1]{\textcolor{black}{#1}}
\newcommand{\sdm}[1]{\textcolor{black}{#1}}
\newcommand{\sd}[1]{\textcolor{black}{#1}}
\newcommand{\scr}[1]{\textcolor{black}{#1}}
\title{Generative Sequential Recommendation with GPTRec}

\author{Aleksandr V. Petrov}
\affiliation{%
  \institution{University of Glasgow} \country{United Kingdom}}

\email{a.petrov.1@research.gla.ac.uk}

\author{Craig Macdonald}
\affiliation{%
  \institution{University of Glasgow} \country{United Kingdom}}
\email{craig.macdonald@glasgow.ac.uk}

\begin{document}
\begin{abstract}
Sequential recommendation is an important recommendation task that aims to predict the next item in \scr{a} sequence. Recently, adaptations of language models, particularly Transformer-based models such as SASRec and BERT4Rec, \scr{have} achieved state-of-the-art results in sequential recommendation. In these models, item ids replace tokens in the original \scr{language} models. However, this approach has limitations. First, the vocabulary of item ids may be many times larger than \sd{in} language models. Second, the classical Top-K recommendation approach used by these models may not be optimal for complex recommendation objectives, including auxiliary objectives such as diversity, coverage or coherence.  Recent progress in generative language models inspires us to revisit generative approaches to address these challenges. This paper presents the GPTRec sequential recommendation model, \scr{which is based} on the GPT-2 architecture. GPTRec can address large vocabulary issues by splitting item ids into sub-id tokens using a novel \se{SVD Tokenisation algorithm} based on quantised item embeddings from \scr{a} SVD decomposition of the user-item interaction matrix. The paper also presents  a novel Next-K recommendation strategy, which generates recommendations item-by-item, considering already recommended items. The Next-K strategy can be used for producing complex interdependent recommendation lists. We experiment with GPTRec on the MovieLens-1M dataset and show that using sub-item tokenisation GPTRec can match the quality of SASRec while reducing the embedding table by 40\%. We also show that \scr{the} recommendations generated by GPTRec on \scr{MovieLens-1M} using the Next-K recommendation strategy match the quality of SASRec in terms of NDCG@10, meaning that the model can serve as a strong starting point for future research. 
\end{abstract}

\maketitle

\section{Introduction} \label{sec:intro}
Sequential recommender systems are \sd{a} class of recommender systems that use the order of user-to-item interactions. These systems are beneficial when the nature of user-item interactions includes sequential patterns. For example, it is natural to watch movies in a particular order (from the first to the last movie in \scr{a} series). Usually, the sequential recommendation task is cast as a next-item prediction task: the goal of the recommender system is to predict the next item in the sequence of user-item interactions. Most of the recent results in sequential recommendation were achieved by deep learning models~\cite{GRU4Rec, Caser, yuanSimpleConvolutionalGenerative2019}, and in particular on Transformer~\cite{Transformer} architecture~\cite{SASRec, BERT4Rec, Bert4RecRepro, PetrovRSS22, DuoRec}. \se{Most of the Transformer-based recommender models, such as SASRec~\cite{SASRec} and BERT4Rec~\cite{BERT4Rec} adapt original language models to recommendation field \scr{by} using item ids instead of tokens \scr{in the language model architecture.} }

The rapid progress of large generative models, such as T5~\cite{raffelExploringLimitsTransfer2020} and the GPT family~\cite{radfordImprovingLanguageUnderstanding, gpt2, brownLanguageModelsAre2020, ouyangTrainingLanguageModels2022} demonstrated that these models could be successfully used for a broad class of tasks, such as text summarisation, sentiment analysis, machine translation etc. \scr{In the recommendation task, these models can be fine-tuned to  directly generate item ids as text strings by a prompt that contains the user's history~\cite{gengRecommendationLanguageProcessing2022}}. \se{Similarly, in a search task, Tay et\ al. showed that generative models can replace \scr{the} traditional reverse index-based retrieval and directly generate relevant document ids by a given query.} This progress motivates us to re-examine the \scr{role} generative \scr{of} models for the sequential recommendation task. This paper discusses some of the challenges in sequential recommendation and how these challenges can be addressed using generative models. We identify and discuss two main challenges \se{and discuss the ramifications of our findings for other related \craig{information retrieval (IR)} tasks}.

First, existing sequential models usually employ a score-and-rank approach to generate recommendations (\se{we also call this the {\em Top-K strategy}}). The problem with this method is that items are scored independently, and similar items are likely to have similar scores. In this case, the model output is likely to be dominated by similar types of items, which may be sub-optimal, and sometimes it is better to show a user different types of items. Instead, we propose a generative \se{\emph{Next-K strategy}}, where recommendations are generated item-by-item. In that case, when the model generates a recommendation in position $K$, it already knows what items are recommended in positions $1..K-1$, and may adjust the result accordingly.  \se{We discuss the details \craig{of these} recommendation generation strategies in Section~\ref{sec:top_k_next_k}.}

Second, a substantial \scr{contribution to the} size of existing \scr{recommendation} models is the item embeddings table.
\se{For example, in a dataset with 10M items and 256-dimensional item embeddings, the embedding table \scr{in a typical recommendation model} would require more than 10GB of GPU memory only for storing embeddings (even without considering gradients and model weights). When we consider gradients, model weights and intermediate scores,  GPU memory requirements increase even more, making model training \se{infeasible}. Realistically, training \scr{state-of-the-art} sequential recommender \scr{models, such as BERT4Rec~\cite{BERT4Rec},} is problematic when there are more than 1M items in the catalogue (see also~\cite{PetrovRSS22})}. We show \se{that this problem can be solved by splitting item} ids into sub-item \scr{\emph{tokens}} \se{using the proposed SVD Tokenisation approach. To obtain sub-item tokens, SVD Tokenisation quantises item embeddings obtained from the SVD decomposition of the user-item interaction matrix.}  Then, the items can be generated token-by-token (similar to how words are generated from sub-word tokens in texts). \se{Section~\ref{ssec:svdtokeniser} covers the details of the SVD Tokenisation algorithm.} 

\looseness -1 To show the feasibility of solving both challenges, we build a generative GPTRec recommendation model based on the GPT-2~\cite{gpt2} architecture. Architecturally, GPTRec is similar to the existing SASRec model~\cite{SASRec}. \sdm{However, GPTRec supports the Next-K recommendation strategy, allowing for more flexible objectives. In addition, GPTRec can work with sub-item tokenisation, which is beneficial for GPU memory consumption. GPTRec also uses a different loss function compared with SASRec} (Cross-Entropy instead of Binary Cross Entropy); empirically, we show that this change is beneficial. We evaluate GPTRec on the MovieLens-1M dataset~\cite{harperMovieLensDatasetsHistory2015} and show the viability of generative techniques for solving the discussed challenges. In particular, \sdm{we observe that \sd{although} we do not specifically tune GPTRec for Next-K generation, it achieves a strong ranking quality similar to SASRec (0.105 NDCG@10 for GPTRec-NextK vs.\ 0.108 SASrec). We also observe that \scr{using} SVD Tokenisation \scr{with} GPTRec can reduce \scr{the} embedding table size by 40\%\scr{,} while achieving comparable ranking quality with SASRec (both have NDCG@10 of 0.108)}.
In short, we summarise the contributions of this paper as follows: 
\begin{enumerate}[topsep=3pt,label={\arabic*.}]
    \item We propose a \sdm{novel} Next-K recommendation strategy as an alternative to traditional Top-K recommendation and \craig{demonstrate its viability by} showing that GPTRec in Next-K generation mode can achieve results similar to those of SASRec.
    \item We propose \sdm{a novel SVD-based} sub-item item tokenisation and token-by-token item generation \scr{approach}. We demonstrate the potential to reduce the embedding table size. We also show that our approach can achieve similar results to SASRec while requiring less storage for embedding tables. 
    \item We propose GPTRec, a generative sequential recommendation model based on the GPT-2 architecture. We show that it achieves results similar to transformer-based models (better than SASRec,  \sdm{and comparable with} BERT4Rec). \sdm{However, in contrast with BERT4Rec, GPTRec is a generative model, \craig{uses SVD Tokenisation for memory efficiency}, and more flexible using the Next-K generation strategy.}
    
    \item{We identify the limitations of proposed approaches and identify strategies to overcome these limitations in future work.}
\end{enumerate}

The rest of the paper is organised as follows: Section~\ref{sec:related} covers existing work related to generative sequential recommendation; Section~\ref{sec:top_k_next_k} discusses limitations of existing Top-K recommendation strategies and proposes Next-K as an alternative; in Section~\ref{sec:gptrec} we introduce GPTRec; Section~\ref{sec:gptvssas} compares GPTRec with existing Transformer-based \scr{models}; Section~\ref{sec:experiments} covers the experimental evaluation of GPTRec; Section~\ref{sec:conclusion} contains directions for further work, \sdm{ramifications for related IR tasks} and concluding remarks. 

\section{Related work} \label{sec:related}
\se{This section covers existing work related to the generative sequential recommendation}: Section~\ref{ssec:related:lms_as_recsys} discusses adaptations of language models for sequential recommendation; Section~\ref{ssec:related:rec_ax_textgen} covers the recent line of work of using pre-trained language models for recommendation task. 
 
\subsection{Adaptations of Language Models for Sequential Recommender Systems} \label{ssec:related:lms_as_recsys}
\looseness -1 Natural Language Processing (NLP) is a prominent and established research field focusing on text-related tasks, such as question answering, machine translation or text summarisation. A text document consists of a sequence of words, and models that work with text have to be able to work with these sequences. 
Sequences of user-item interactions in sequential recommender systems have a very similar structure to the sequences of words in texts. Therefore recommender systems researchers frequently adapt models originally developed for language processing to the sequential recommendation task. For example, GRU4Rec~\cite{GRU4Rec}, one of the first neural architectures for the sequential recommendation, is based on the GRU model~\cite{choPropertiesNeuralMachine2014}, originally designed for machine translation tasks. 

Recently, many \se{recommendation} models have adopted variations of \craig{the} Transformer~\cite{Transformer} architecture, which was also originally designed for \craig{addressing} machine translation. For example, two of the most popular Transformer-based sequential recommenders are (i) SASRec~\cite{SASRec}, which uses the decoder part of the Transformer model and is trained to shift the input sequence of items one element to the left (in language models, this task is also known as the Language Modelling task, \sd{LM}) and (ii) BERT4Rec~\cite{BERT4Rec}, which, in contrast, is trained to recover masked items from the sequence (this is known as Masked Language Modelling, MLM).

Other more recent Transformer-based sequential models include DuoRec~\cite{DuoRec}, CBiT~\cite{CBiT}, ALBERT4Rec~\cite{Bert4RecRepro} and many others. While these works slightly change the architecture or augment the model with additional training tasks, they all share a Transformer encoder or a Transformer decoder as the main component of their architecture. \se{Despite achieving state-of-the-art results in ranking metrics, such as NDCG@K, the Top-K recommendation strategy employed by these models is not flexible enough to optimise for other metrics such as diversity or serendipity; these models also suffer from high GPU requirements when the number of items in the catalogue is high. \sdm{These problems and the recent \scr{advances} in generative models motivate us to apply \scr{a} generative approach for sequential recommendation.}}

\subsection{Recommendations as Text Generation}\label{ssec:related:rec_ax_textgen}
The arrival of Large Language Models, such as T5~\cite{raffelExploringLimitsTransfer2020} and GPT-3~\cite{brownLanguageModelsAre2020}, has shown that the pre-trained text generation models may serve as universal problem solvers and can be applied to a large class of tasks. 

This was specifically shown for recommendation in recent works. For example, the P5 model~\cite{gengRecommendationLanguageProcessing2022} uses text generation to generate item and user ids directly as text strings. The M6-Rec~\cite{cuiM6RecGenerativePretrained2022} model directly generates item titles as recommendations. TIGER~\cite{rajputshashankRecommenderSystemsGenerative} introduces the idea of \emph{semantic id}, where the item is represented as a set of tokens derived from its side information, such as product category.
While using pre-trained models for recommendation is an interesting research direction, it differs from ours. Indeed, these models rely on the existence of pre-trained models, which encapsulate knowledge about the world (including the knowledge about the recommended domain), and, therefore, can be seen as recommender systems with side information.  
In contrast, we do not rely on any side information and research \scr{a} more classical sequential recommendation setting \scr{where} the only data available to the model is the interaction sequences.  This allows us to understand the properties of the generative approach better and decouple it from the benefits of the availability of side information.

\se{While this paper focuses on the sequential recommendation task, the discussed problems are relevant for a broader field of Generative IR.
Indeed, in the Generative IR \sdm{tasks, which directly generate document ids (docids) for a given query}, a large number of docids (the equivalent of items)  and the techniques for splitting atomic document ids into tokens are frequently mentioned  as one of the central problems in the recent Generative IR publications~\cite{ zhouDynamicRetrieverPretrainedModelbased,tayTransformerMemoryDifferentiable}. \sdm{\sd{Some of the recent papers have specifically focused on sub-docid tokenisation~\cite{sunLearningTokenizeGenerative2023, mehtaDSIUpdatingTransformer2022} and inspire us to develop a similar technique for recommender systems}}. SVD-based tokenisation (or, more broadly, embedding-based tokenisation) described in this paper can be adapted to the IR domain and help to solve scaling issues in the IR. Complex interdependent search results are also an active area of IR research,  including, for example, results diversity~\cite{xiaLearningMaximalMarginal2015} and fairness~\cite{jaenichColBERTFairPRFFairPseudoRelevance2023}. The Next-K generation strategy proposed in this paper can help to resolve these issues.}

\se{In \sdm{summary, existing Transformer-based 
 sequential recommendation} models achieve high quality in ranking metrics, but they are inflexible to optimise for other metrics and have high GPU memory requirements. On the other hand, recent works show the potential of pre-trained language models for the sequential recommendation problem. However, they rely on side information and pre-existing knowledge, which is not always available or reliable.} \se{To address the objective flexibility problem, in the next section, we propose \craig{a} Next-K recommendation strategy as an alternative to Top-K, \sdm{specifically designed to work well with  generative models}. We then describe GPTRec in Section~\ref{sec:gptrec}, which can utilise the Next-K strategy and may use SVD Tokenisation to reduce GPU memory consumption.}

\section{Top-K and Next-K recommendations}\label{sec:top_k_next_k}
In this section, we discuss the classical Top-K recommendation approach \se{(used by BERT4Rec and SASRec)} and its fundamental limitations and propose a novel alternative Next-K approach for generative models. 

\begin{algorithm}[t]
\caption{Top-K Recommendation Strategy}
\label{alg:top_k}
\begin{algorithmic}[1]
    \Require{$user$ is the target user for recommendations}
    \Require{$items$ is the set of all available items}
    \Require{$K$ is the number of recommendations to generate}
    \Require{$f$ is an existing scoring model}
    \Procedure{TopKRecommend}{$user, items, K, f$}
        \State $item\_scores \gets \text{empty list}$
        \For{$item \in items$}
            \State $score \gets f(user, item)$
            \State $item\_scores.\text{append}((item, score))$
        \EndFor
        \State $sorted\_items \gets \text{sorted}(item\_scores,$ \\
        \hspace*{1.3cm} $\text{key}=\text{score}, \text{reverse}=\text{True})$
        \State $recommended\_items \gets [item$ \\
        \hspace*{1.3cm} $\text{ for } (item, score) \text{ in } sorted\_items[:K]]$
        \State \textbf{return} $recommended\_items$
    \EndProcedure
\end{algorithmic}
\end{algorithm}

\subsection{Top-K strategy} \label{ssec:topk}
The ultimate goal of a recommender system is to provide a user with a list of items which are likely to be of user interest. The most typical approach for solving this is the \emph{Top-K} recommendation strategy~\cite{leeDifferentiableRankingMetric2021,zhuImprovingTopKRecommendation2019}. In this strategy, the recommendation model $f(user, item) \rightarrow s$ returns an \craig{estimated relevance} score $s$ for \se{each} $(user, item)$ pair (in the case of sequential recommendation, $user$ is actually represented as a sequence of interactions). To generate the recommendation for a particular user $u$, the recommender system uses model $f$ to generate all item relevance scores and then selects $k$ items with the highest scores as the recommendations. Algorithm~\ref{alg:top_k} illustrates this strategy using pseudo-code. 

\looseness -1 The fundamental problem with the Top-K strategy is that it assumes that every item can be scored independently from the other recommended items.
\se{However, the Top-K strategy fails when the recommendations should be interdependent.} For example, there is a redundancy problem: if a user has recently bought a coffee machine, they will likely buy coffee beans as their next purchase. Therefore, it is a good strategy to recommend coffee beans to the user. However, there can be many different variants of coffee beans on sale. Each variant will likely have a high recommendation score, so most recommended items will be coffee beans. \craig{This} leads to redundant recommendations (it is enough to recommend only 1-2 types of coffee) and poor representation of the user's other interests~\scr{\cite{zhou2009accurate}}.
\se{Diversification re-ranking methods, such as MMR~\cite{carbonellUseMMRDiversitybased1998}, and multi-aspect methods~\cite{caiAspectRedistributionLearning2022} can address redundancy and diversity, but they pose challenges such as increased complexity, computational demands, and difficulty in training. Additionally, these approaches may require extensive hyperparameter tuning to achieve optimal performance~\scr{\cite{sener2018multi}}. Instead of re-ranking approaches, training a single model with an interdependent objective can provide a more efficient solution. }

Other problems that Top-K recommender systems will likely fail to solve include complementary item recommendations (e.g., we may want to recommend fashion items which fit well with each other), serendipity (sometimes we want to help users discover new items, but only if we already recommended items of their interest), and coverage (we may want to be sure that our recommendations provide good coverage of the system's catalogue) etc. We now introduce the Next-K recommendation strategy, \se{capable} to resolve these problems.

\begin{algorithm}
\caption{Next-K Recommendation Strategy}
\label{alg:next_k}
\begin{algorithmic}[1]
\Require{$user$ is the target user for recommendations}
\Require{$items$ is the set of all available items}
\Require{$K$ is the number of recommendations to generate}
\Require{$f$ is an existing scoring model}
\Procedure{NextKRecommend}{$user, items, K, f$}
    \State $recommended\_items \gets \text{empty list}$
    \For{$k = 1$ to $K$}
        \State $max\_score \gets -\infty$
        \State $best\_item \gets \text{None}$
        \For{$item \in items \setminus recommended\_items$}
            \State $score \gets f(user, recommended\_items, item)$
            \If{$score > max\_score$}
                \State $max\_score \gets score$
                \State $best\_item \gets item$
            \EndIf
        \EndFor
        \State $recommended\_items.\text{append}(best\_item)$
    \EndFor
    \State \textbf{return} $recommended\_items$
\EndProcedure
\end{algorithmic}
\end{algorithm}

\subsection{Next-K strategy} \label{ssec:next-k}
Next-K is the recommendation strategy, where the recommender system decides what to recommend at position $k$ only after generating recommendations at position $1 .. k-1$. 
More formally, the item score depends not only on the user and item but also recommendations on previous positions: $f(user, {recommended\_items}, item) \rightarrow score$.
The model iteratively scores all items (excluding already recommended) to generate recommendations and adds the highest-scored item to the $recommended\_items$ list.  Algorithm~\ref{alg:next_k} illustrates the strategy in pseudo-code. 
Because the Next-K strategy considers already recommended items, it may address the issues of the classic Top-K strategy, such as lack of diversity and poor \se{representation} of user interests. 

A limitation of the Next-K strategy is that it is more computationally expensive: it requires the generation of the full scores distribution $k$ times for a user, whereas the Top-K strategy needs only one inference per user. Another problem is that the scoring function now has more \scr{input} \scr{arguments} \scr{(i.e. in addition to the user id  and the item, it takes into account already recommended items)}, so training such a function may be challenging.    

In summary, in this section, we described a popular Top-K  recommendation strategy and identified the limitations of this strategy. We also proposed the Next-K strategy as a solution to the Top-K strategy problems and also identified the limitations of this strategy. In the next section, we introduce the GPTRec model, which can be used with both Top-K and Next-K strategies. \se{We now turn to GPTRec, a generative recommendation model that can use either Top-K or Next-K recommendation strategies.}

\section{GPTR\protect\lowercase{ec}}
 \label{sec:gptrec}
In this section, we introduce GPTRec, a generative sequential recommendation model  we use as a backbone for our experiments. 

\begin{table*}
    \caption{GPU memory requirements for the multi-token-per-item model, as a percentage of the one-token-per-item model, for selected datasets. $t$ corresponds to the number of tokens per item, and $v$ corresponds to the number of possible items per token in the SVD Tokenisation algorithm. The number of embeddings and GPU memory requirements in each multi-token-per-item configuration is shown in the brackets. We assume that each embedding is stored as 256 float32 numbers.}\label{tb:memory_reduction}
    \resizebox{\textwidth}{!}{
        \sdm{
            \begin{tabular}{lll|ccc|ccc|ccc}
\toprule
Dataset & Num Items & \makecell[l]{GPU Memory \\ (one-token-per-item \\ mode)} & 
\makecell{t=2 \\ v=128 \\ (256 embs;\\ 0.25 MB)} & 
\makecell{t=2 \\ v=512 \\ (1024 embs;\\ 1.00 MB)} & 
\makecell{t=2 \\ v=2048 \\ (4096 embs;\\ 4.00 MB)} & 
\makecell{t=4 \\ v=128 \\ (512 embs;\\ 0.50 MB)} & 
\makecell{t=4 \\ v=512 \\ (2048 embs;\\ 2.00 MB)} & 
\makecell{t=4 \\ v=2048 \\ (8192 embs;\\ 8.00 MB)} & 
\makecell{t=8 \\ v=128 \\ (1024 embs;\\ 1.00 MB)} & 
\makecell{t=8 \\ v=512 \\ (4096 embs;\\ 4.00 MB)} & 
\makecell{t=8 \\ v=2048 \\ (16384 embs;\\ 16.00 MB)} \\
\midrule
MovieLens-1M & 3,416 & 3.34 MB & 7.494\% & 29.977\% & 119.906\% & 14.988\% & 59.953\% & 239.813\% & 29.977\% & 119.906\% & 479.625\% \\
MovieLens-20M & 138,493 & 135.25 MB & 0.185\% & 0.739\% & 2.958\% & 0.370\% & 1.479\% & 5.915\% & 0.739\% & 2.958\% & 11.830\% \\
Yelp & 150,346 & 146.82 MB & 0.170\% & 0.681\% & 2.724\% & 0.341\% & 1.362\% & 5.449\% & 0.681\% & 2.724\% & 10.898\% \\
Gowalla & 1,280,969 & 1.22 GB & 0.020\% & 0.080\% & 0.320\% & 0.040\% & 0.160\% & 0.640\% & 0.080\% & 0.320\% & 1.279\% \\
Amazon Books & 5,264,307 & 5.02 GB & 0.005\% & 0.019\% & 0.078\% & 0.010\% & 0.039\% & 0.156\% & 0.019\% & 0.078\% & 0.311\% \\
LastFM-1b & 32,291,134 & 30.80 GB & 0.001\% & 0.003\% & 0.013\% & 0.002\% & 0.006\% & 0.025\% & 0.003\% & 0.013\% & 0.051\% \\
\bottomrule
\end{tabular}

        }
    }
\end{table*}
\subsection{Architecture}
GPTRec,\footnote{GPT stands for Generative Pre-trained Transformer. In fact, in our adaptation, we don't use the pre-trained versions of GPT, but we saved the letter P in the model name in order to give credit to the GPT authors.} designed for the generative sequential recommendation, utilises the GPT-2 architecture~\cite{gpt2}, which in turn is based on the decoder part of the Transformer model~\cite{Transformer}. There are minor modifications to the original Transformer model in GPT-2, such as:
\begin{itemize}
    \item Moving the layer normalisation to the beginning of the transformer block;
    \item \se{Implementing \scr{a} modified initialisation with scaled residual weights;}
    \item Employing learnable positional encodings instead of sine-based encodings.
\end{itemize}
For brevity, we omit the details of the Transformer model and refer readers to the original publications~\cite{gpt2, Transformer}.

\subsection{Tokenisation} \label{ssec:svdtokeniser}

Similarly to existing Transformer-based sequential recommendation models such as SASRec~\cite{SASRec} and BERT4Rec~\cite{BERT4Rec}, GPTRec can use item ids instead of tokens in the original GPT model. We call this \emph{token-per-item mode}. However, \se{as discussed in Section~\ref{sec:intro}}, when the number of items in the catalogue is large, this approach leads to a massive embedding table, which is hard to fit into a single GPU.

\se{Language models, such as BERT~\cite{BERT} and GPT-2~\cite{gpt2} solve a similar problem} by splitting whole words into sub-word tokens. Example strategies for sub-word tokenisation include Word Piece encoding~\cite{wuGoogleNeuralMachine2016} (used by BERT) and BytePair encoding~\cite{sennrichNeuralMachineTranslation2016} (used by GPT family). Inspired by this idea,  GPTRec can also split items into sub-items to reduce memory footprint. In this case, each item is represented using $t$ tokens, where each of the $t$ tokens can be chosen from $v$ alternatives. We call this \emph{multi-token-per-item mode}. 

\begin{figure*}
    \includegraphics[width=0.8\textwidth]{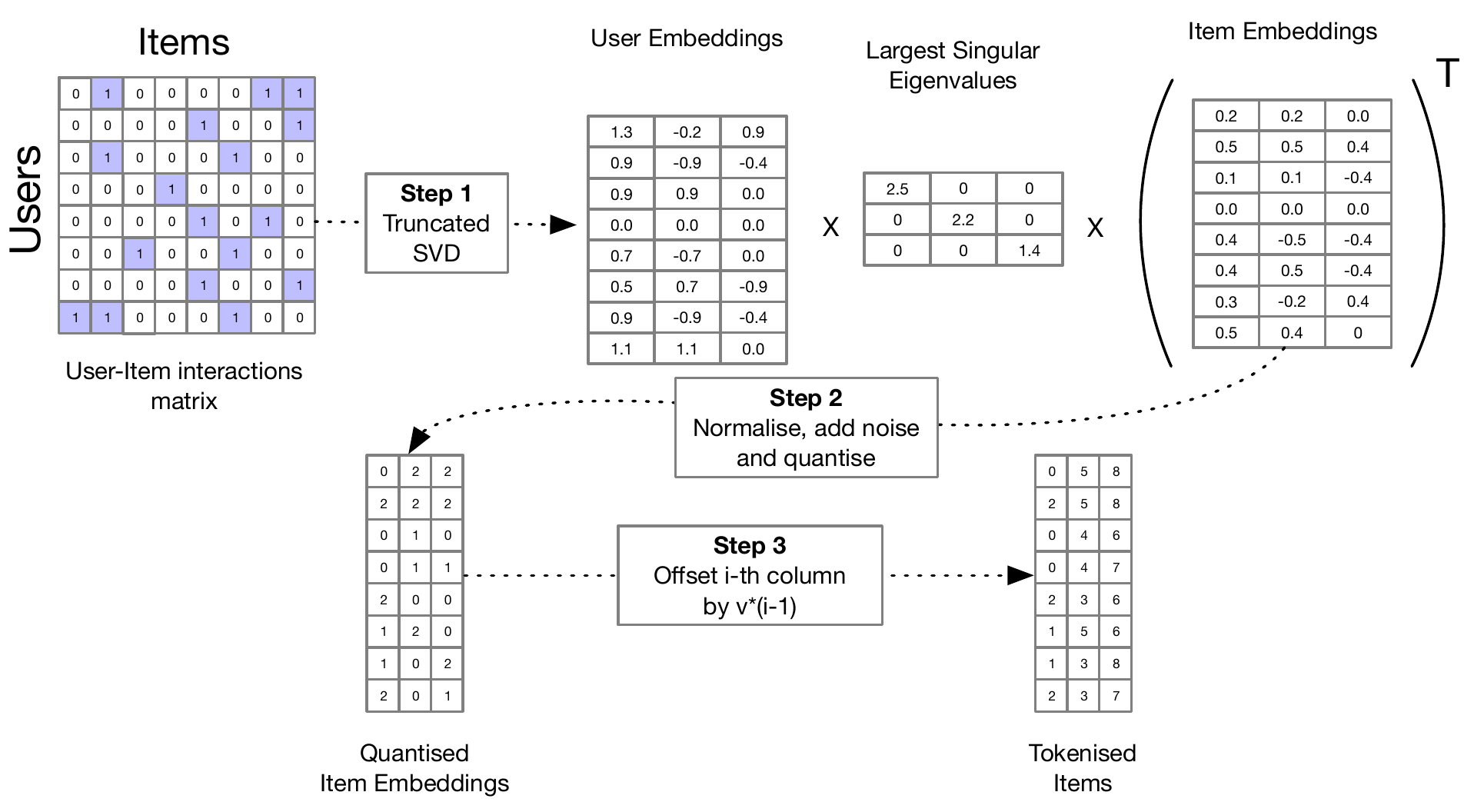}
    \caption{Step-by-step example of the SVD Tokenisation algorithm}
    \label{fig:svdtokenise}
\end{figure*}

As a simple heuristic to decompose items into sub-item tokens, we propose an \se{SVD-based Tokenisation algorithm}. \se{Figure~\ref{fig:svdtokenise} illustrates the main three steps of the algorithm}. 

\se{First (Step 1 of the figure),} the SVD Tokenisation algorithm builds a user-item interaction matrix $M$ and performs a truncated SVD decomposition of this matrix with $t$ latent components: 
\begin{align}
    M \approx U \times \Sigma \times E^{T}
\end{align}
where $U$ is the matrix of user embeddings, $E$ is the matrix of item embeddings, and $\Sigma$ is the diagonal matrix with $t$ largest singular eigenvalues on the diagonal. Both user and item embeddings have $t$ latent components. 

\looseness -1 \craig{In \se{Step 2 of Figure~\ref{fig:svdtokenise}}, the item embeddings $E$ are  normalised,} so that each embedding dimension is in the [0..1] interval \se{and} a small amount of Gaussian noise \sd{is added} to ensure that no two items have equal \craig{embeddings} (equal embeddings may happen if \se{exactly the same users interacted with two items}). In our experiments, we use Gaussian noise with zero mean and \scr{a} standard deviation of $10^{-5}$, which is several orders of magnitude less than the scale of the \craig{embedding} values themselves (they are in the [0..1] range after normalisation). After that, SVD Tokenisation quantises each dimension of item embeddings into $v$ values; each value in the quantised embeddings represents one token in the final representation of the item. 

\se{Finally, (Step 3 of Figure~\ref{fig:svdtokenise}) the algorithm offsets $i^{th}$ dimension of the quantised embeddings by $v*(i-1)$ to}  ensure that every dimension of the item representation has its range of tokens (i.e.\ the first dimension is represented by tokens $[0~..~v-1]$, the second dimension is represented by tokens $[v~..~2v-1]$ and so on. %

Algorithm~\ref{alg:svdtokenizer} illustrates this procedure in pseudo-code. In this algorithm, tokenised items are, in fact, discretised embeddings from \se{an SVD matrix decomposition of the user-matrix interactions matrix}. Similar items are likely to have similar embeddings and, therefore, similar multi-token representations. As for memory consumption, the model now only needs to store $t \times v$ embeddings \se{rather than storing one embedding per item}. \scr{Therefore,} the amount of required GPU memory for storing embeddings does not depend on the total \scr{number} of items in the catalogue; this is similar to how fixed-size token vocabulary can represent millions of different words by combining different tokens in language models. \se{For example, as we mentioned earlier, storing the embedding table in the one-token-per-item mode for 10M items requires more than 10GB of GPU memory. However, embedding a table for the same dataset in multi-token-per-item mode with $t=8$ and $v=2048$ requires only 16MB GPU Memory (only 0.16\% of the original embedding table size)}. 
\se{Table~\ref{tb:memory_reduction} illustrates similar calculations for several available recommender systems datasets and different configurations of the SVD Tokenisation algorithm. }

\begin{algorithm}[tb]
\caption{SVD Tokenisation Algorithm}
\label{alg:svdtokenizer}
\begin{algorithmic}[1]
    \Require{$M$ is the user-item interaction matrix}
    \Require{$t$ is the number of tokens per item}
    \Require{$v$ is the number of alternatives per token}
    \Ensure{Sub-item token representation of items}
    \Procedure{SVDTokenise}{$M, t, v$}
        \State Compute item embeddings $E$ with truncated SVD on $M$
        \For{$i = 1$ to $t$}
            \State \scr{Normalise} $E_i$ to range $[0, 1]$
            \State Add small Gaussian noise to $E_i$
            \State Quantise $E_i$ into $v$ bins
            \State Offset quantised values by $(i - 1) * v$
        \EndFor
        \State Concatenate quantised components for sub-item tokens
        \State \textbf{return} Sub-item token representation
    \EndProcedure
\end{algorithmic}
\end{algorithm}

\subsection{Training Objective and Loss Function}
\label{ssec:objective_and_loss}
Following GPT-2~\cite{gpt2}, we train the model using the Language\footnote{Even though our models do not work with texts, we retain the "Language" in the name of the objective to stay consistent with the NLP literature that serves as a source of inspiration for us.  } Modelling objective with Cross-Entropy loss. The language modelling approach factorises the probability of a sequence of tokens $S=\{s_1, s_2, ... s_n\}$ (recall that in GPTRec, $s_{i}$ represents an item or a sub-item token) as the product of conditional probabilities:
\begin{align}
    p(S) = \prod_{i=1}^n p(s_i|s_1, s_2, ... s_{i-1})\label{eq:prob-factor}
\end{align}
The loss function is then derived from Equation~\eqref{eq:prob-factor} using the Maximum Log-likelihood principle~\cite[Ch.5]{goodfellowDeepLearning2016}:
\begin{align}
    \mathcal{L}_{LM} =-\frac{1}{n}\sum_{i=1}^{n}\log(p(s_i|s_1, s_2, ... s_{i-1}))
\end{align}

Similarly to GPT-2, in GPTRec, $p(s_i|s_1, s_2, ... s_{i-1})$ is modelled as $\text{softmax}(\cdot)$ operation over $i^{th}$ output of the model, so overall the model is trained to shift its input one token to the left. The last token predicted by the model is a new token that did not appear in the input sequence.  \se{This shifted sequence can be fed back into the model, which enables us to implement the Next-K recommendation approach for increased flexibility and utilise the multi-token-per-item mode for improved GPU efficiency.}

In short, GPTRec is trained with standard Language modelling (LM) training objective,  \scr{however, instead of token ids},  we use item ids or sub-id values obtained from quantised item embeddings. We now turn to the generation of recommendations. 

\subsection{Generating Recommendations with GPTRec}
In this section, we describe possible options to generate a list of recommendations using GPTRec. 

\subsubsection{Top-K generation, one-token-per-item}
GPTRec supports the Top-K recommendation strategy. 
In the simplest case, when one item corresponds to one token, the generation of recommendations is similar to that of SASRec. GPTRec outputs the probability distribution \se{$p(s_i|s_1, s_2, .. s_{i-1})$} of the next token for each position in the sequence, meaning that the last probability distribution corresponds to the next most likely item in the sequence. In this generation mode, GPTRec uses this last probability distribution as item scores and applies standard Top-K scoring (see Section~\ref{ssec:topk} and Algorithm~\ref{alg:top_k}).

\subsubsection{Top-K generation, multi-token-per-item}
The next generation strategy supported by GPTRec is \emph{Top-K generation with multi-token-per-item}. In this case, the model outputs $K$ candidate items using standard GPT-2 autoregressive generation: each time, the model predicts the probability distribution of the next token and then samples a token from the distribution. This token is added to the end of the input sequence, and the procedure repeats until the model generates $t$ tokens, where $t$ is the number of tokens in each item representation. After generating the candidate, the model scores the candidate using the chain rule described by Equation~\eqref{eq:prob-factor}. Finally, the model applies the standard Top-K strategy Algorithm~\ref{alg:top_k} to the set of items, assuming that any item that was not generated as a candidate has $-\infty$ score.

Note that using this procedure, some generated sequences may not correspond to any \craig{valid} item id; the model ignores these invalid sequences. Some of the generated candidates may also repeat. In practice, this means that the number of candidates $K$ that has to be chosen is relatively high compared to the required number of recommended items; For example, we use $K=50$ even though we only needed ten items to generate recommendations.

\looseness -1 \subsubsection{Next-K generation, one-token-per-item}\label{ssec:next_k}
In this case, the recommendation list is generated using the Next-K procedure (Algorithm~\ref{alg:next_k}). At each iteration, the model concatenates the sequence of the user's interactions with the sequence of already generated recommendations and generates the next most likely item. \scr{Next-K strategy is also equivalent to the \emph{greedy search} token generation strategy in language models.\footnote{\scr{To use greedy search (the equivalent of the  Next-K strategy) in HuggingFace Transformers library,  one can call the \texttt{.generate()} method with parameters  \texttt{num\_beams=1} and \texttt{do\_sample=False}.\\ See also   \scriptsize{\href{https://huggingface.co/docs/transformers/v4.30.0/en/generation\_strategies\#greedy-search}{https://huggingface.co/docs/transformers/v4.30.0/en/generation\_strategies\#greedy-search}}.}}}

Note that the training objective, in this case, is somewhat misaligned with the generation procedure. Indeed, the model is trained to predict one next token and not $K$ next tokens. Therefore we expect quality degradation with increased $K$. To align the training procedure with the Next-K generation procedure, the model requires more advanced training objectives, such as reinforcement learning techniques described in the InstructGPT~\cite{ouyangTrainingLanguageModels2022} paper.
Nevertheless, these techniques require a model that can already generate some plausible sequences. Therefore, \craig{in this paper, we examine} if a model trained for Top-K generation can return meaningful recommendations in the Next-K mode, \craig{and leave} reinforcement learning fine-tuning for future work.    
This concludes the description of the GPTRec model. \se{In the next section, we compare GPTRec with existing SASRec and BERT4Rec models} before moving to experimental analysis of GPTRec in Section~\ref{sec:experiments}.

\section{Comparison of GPTR{\protect\lowercase{ec}} with SASR{\protect\lowercase{ec}} and BERT4R\protect\lowercase{ec}}\label{sec:gptvssas}
In this section, we compare GPTRec with the two most popular transformer-based sequential models, namely with SASRec~\cite{SASRec} and BERT4Rec~\cite{BERT4Rec}.
SASRec~\cite{SASRec} is the most similar model to GPTRec. Indeed, both models use the decoder part of the Transformer~\cite{Transformer} architecture and are trained to shift the input sequence one element to the left.
However, GPTRec differs from SASRec by using the Cross-Entropy loss (\se{aka Softmax Loss}), while SASRec is trained using Binary Cross Entropy loss. The use of Cross-Entropy loss is directly derived from the Maximum Log likelihood principle. \se{In contrast, Binary Cross-Entropy used by SASRec is a heuristic, which, as we show in Section~\ref{sec:experiments}, underperforms compared to Cross-Entropy loss}. BERT4Rec, on the other hand, uses Cross-Entropy loss, but it uses a Masked Language Modelling task (MLM; also known as Cloze, or Item Masking), which is less aligned with the next item prediction task~\cite{Bert4RecRepro}. BERT4Rec also differs from GPTRec by using the encoder part of the Transformer, whereas SASRec and GPTRec use the decoder part.

Second, SASRec only supports Top-K recommendations in one-token-per-item mode. In \se{contrast, GPTRec also} supports a multi-token-per-item mode to reduce memory footprint and a Next-K recommendation strategy, which can be used for learning more advanced recommendation objectives using techniques such as reinforcement learning. 

\se{In summary, compared with existing models, GPTRec is more GPU-memory efficient when trained in multi-token-per-item mode. It also allows for more flexible recommendation generation using the Next-K generation strategy.} The left side of Table~\ref{tb:transformers_comparison} also portrays the salient characteristics of these models. In the next section, we evaluate GPTRec experimentally, compare it with SASRec and BERT4Rec, and experimentally study its performance in multi-token-per-item and Next-K generation modes. 

\section{Experiments}\label{sec:experiments}
This section covers the experimental evaluation of GPTRec. We aim to answer the following research questions: 

\begin{enumerate}[font={\bfseries}, label={RQ\arabic*}, wide, labelwidth=!, labelindent=0pt]
    \item How does GPTRec perform in one-token-per-id mode compared to BERT4Rec and SASRec? 
    \item What are the effects of the number of tokens and  the number of values per token in the multi-token-per-item mode in GPTRec? 
    \item What is the effect of cutoff level $K$ on the GPTRec performance in the Next-K recommendation mode? 
\end{enumerate}

\subsection{Experimental Setup}
We implement GPTRec using the GPT-2 architecture from HuggingFace Transformers~\cite{wolfHuggingFaceTransformersStateoftheart2020} and the aprec\footnote{\href{https://github.com/asash/bert4rec\_repro}{https://github.com/asash/bert4rec\_repro}} framework from a recent replicability study~\cite{Bert4RecRepro}. We follow the experimental setup described in the same replicability study~\cite{Bert4RecRepro} and use SASRec's and BERT4Rec's results reported in the paper as baselines.

We use the MovieLens-1M dataset~\cite{harperMovieLensDatasetsHistory2015} for the experiments. While there are some known issues with the dataset (e.g.\ users don't rate movies in the dataset in the same order as they watch them), the dataset is used consistently across many papers, allowing compare the published results across papers. The salient characteristics of the dataset are listed in Table~\ref{tb:dataset}.

\se{We employ a leave-one-out strategy for model testing, holding out each user's latest interaction in the test set. In addition, for 128 randomly selected users, we hold out their second-to-last action in a separate validation set. This validation set is used for the early stopping mechanism and to control model quality during training, ensuring that our model generalises well to new data.}

Following common practice~\cite{SASRec, BERT4Rec, Bert4RecRepro} we use a relatively shallow transformer model with three transformer blocks. We set the maximum sequence length to 100 and truncate the sequence to the \craig{last} 100 interactions if the user interacted with more items. We use 256-dimensional token embeddings in our experiments. 
We use an early \scr{stopping} mechanism \scr{that terminates training} if \scr{validation  } NDCG@10 does not improve for 300 epochs. 

\begin{table}
    \caption{Salient characteristics of the MovieLens-1M\ dataset}\label{tb:dataset}
    \begin{tabular}{c|c}
    \toprule
     Number of users & 6040 \\
     Number of items & 3416 \\
     Number of interactions & 999611 \\
     Average sequence length & 165.49 \\
     Median sequence length & 96 \\
     \bottomrule
    \end{tabular}
\end{table}

\subsection{RQ1. Comparison with existing transformer-based models}
\begin{table*}
    \caption{Model performances on the MovieLens-1M dataset.
        \textbf{Bold} values indicate the best results, and \underline{underlined} values indicate the second best results. SASRec and BERT4Rec results are copied from the reproducibility paper~\cite{Bert4RecRepro}.}
        \label{tb:transformers_comparison}
    \resizebox{\textwidth}{!}{
        \begin{tabular}{lllll|ll}
\toprule
Model name & Generation Strategy & Architecture & Training Task & Loss & Recall@10 & NDCG@10 \\
\midrule
BERT4Rec & TopK & Encoder & MLM (Item Masking) & Cross Entropy (Softmax Loss) & \textbf{0.282} & \textbf{0.152} \\
GPTRec-TopK & TopK & Decoder & LM (Sequence Shifting) & Cross Entropy (Softmax Loss) & \underline{0.254} & \underline{0.146} \\
SASRec & TopK & Decoder & LM (Sequence Shifting) & Binary Cross-Entropy & 0.199 & 0.108 \\
GPTRec-NextK & NextK & Decoder & LM (Sequence Shifting) & Cross Entropy (Softmax Loss) & 0.157 & 0.105 \\
\bottomrule
\end{tabular}

    }

\end{table*}

We first evaluate the performance of GPTRec in comparison with popular Transformer-based sequential recommendation models. For this comparison, we choose BERT4Rec~\cite{BERT4Rec} and SASRec~\cite{SASRec} as the baselines and compare them with two variations of the GPTRec model: GPTRec-TopK (GPTRec with the Top-K recommendation generation strategy) and GPTRec-NextK (GPTRec with the Next-K generation strategy). \se{In this experiment, we use all models, including GPTRec,} in the one-token-per-item mode. Table~\ref{tb:transformers_comparison} reports the result of our comparison. 

The table shows that the results achieved by GPTRec-TopK are \sd{similar to} the results of SASRec and BERT4Rec: e.g.\ it achieves NDCG@10 of 0.146, which is better than SASRec's result (0.108, +35\%) but \sdm{comparable} to BERT4Rec's result (0.152, -4\%). The results measured by the Recall@10 metric also follow the same pattern: GPTRec achieves Recall@10 of 0.254, \sdm{slightly} worse than BERT4Rec's result (0.282) but better than SASRec's result (0.199). 

Overall we can say that the results of GPTRec-TopK are comparable with BERT4Rec and better than SASRec. 

The \craig{observation} that GPTRec-TopK performs better than SASRec is specifically interesting. As discussed in Section~\ref{sec:gptvssas}, when GPTRec is used in one-token-per-item mode with Top-K generation strategy, its main difference with SASRec is the loss function (Cross-Entropy in GPTRec vs.\ Binary Cross-Entropy in SASRec). This confirms that Cross-Entropy loss is better for the next-item recommendation. 

From Table~\ref{tb:transformers_comparison}, we also see that GPTRec-NextK performs worse than other models. This is not surprising: in Section~\ref{ssec:next_k}, we discussed that the model would likely require more complex tuning techniques, such as reinforcement learning, to perform well in the Next-K generation mode. However, as \craig{can be seen} from the table, \craig{the} NDCG@10 achieved by the model is very similar to SASRec's, which means that the model can serve as a strong starting point for further tuning, \scr{for example  using reinforcement learning\footnote{\scr{Reinforcement learning is needed because, in the Next-K strategy, we can only compute our utility function (e.g.\ NDCG) after conducting inference using the model K times. Therefore, it is hard to optimise the model using classic gradient descent, which assumes that it is possible to compute the loss function after every model inference.}}}. 

Overall in answer to RQ1, we conclude that GPTRec-TopK achieves similar results to BERT4Rec and outperforms SASRec. On the other hand, GPTRec-NextK performs worse than other models but is still comparable with SASRec's results and may serve as a starting point for future research.

\subsection{RQ2. Effect of the number of tokens and the number of tokens per item in multi-token-per-item mode} 
\begin{figure*}
    \subfloat[Recall@10]{
        \includegraphics[width=0.4\textwidth]{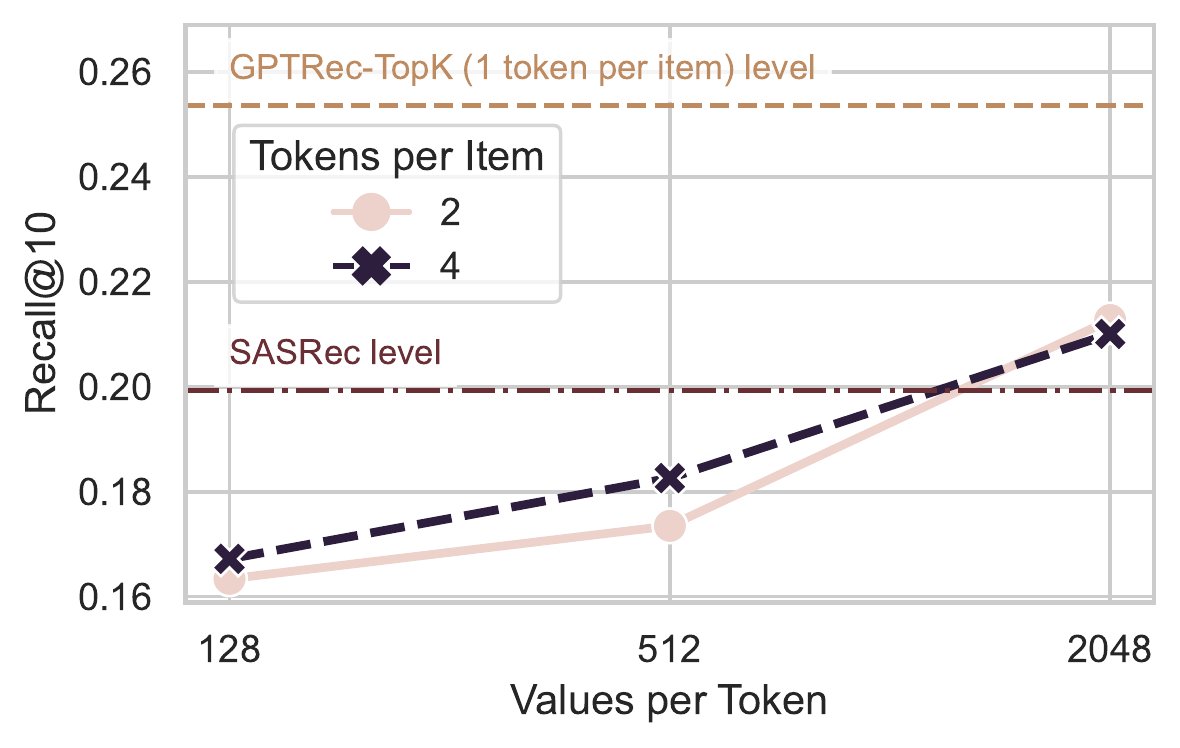}}{
        \label{fig:recall10_tokens}
    }\subfloat[NDCG@10]{
        \includegraphics[width=0.4\textwidth]{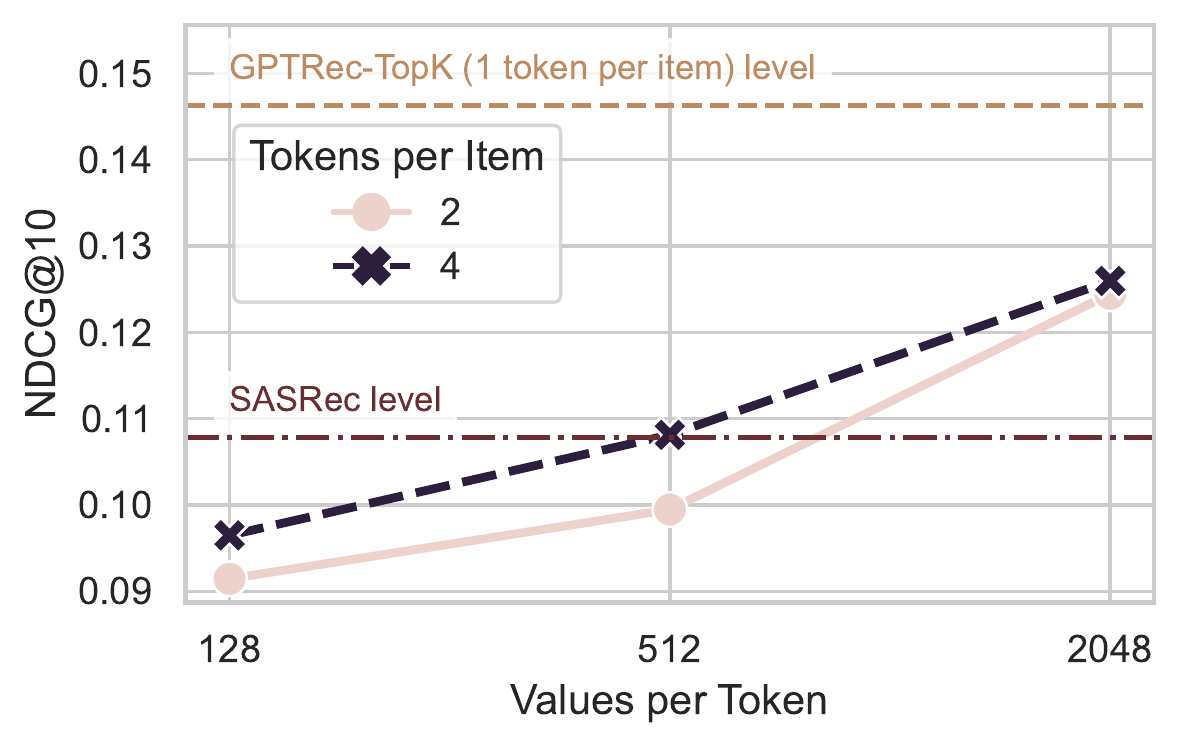}}{
        \label{fig:ndcg10_tokens}
    }
    \caption{GPTRec performance in multi-token-per-item mode. The figure also includes the performance of SASRec for comparison, as well as 
    GPTRec in one-token-per-item mode.} \label{fig:multi-token}
\end{figure*}

To answer our second research question, we train GPTRec in multi-token-per-item mode. To split item ids into sub-item tokens, we use \se{the SVD Tokenisation algorithm}, as described in Section~\ref{ssec:svdtokeniser}. 
We select the number of tokens per item from \sdm{\{2, 4\}} and the number of values per token from \sdm{\{128, 512, 2048\}}. 

Figure~\ref{fig:multi-token} \craig{provides} the \se{Recall@10 and NDCG@10 metrics} of these models and compares them with GPTRec-TopK trained in one-token-per-item mode and SASRec. 
The figure shows that the model's performance degrades when compared to GPTRec-TopK in  one-token-per-item mode but remains competitive with SASRec. For example, with four tokens per item and 512 values per token GPTRec almost exactly matches the performance of SASRec in terms of NDCG@10 (0.108 for both models) but is worse when measured on Recall@10 (0.182 for GPTRec with four tokens and 0.199 for SASRec). However, this is substantially worse than GPTRec-TopK in one-token-per-item, which achieves NDCG@10 of 0.253. On the other hand, multi-token-per-item mode uses less GPU memory. For example, with four tokens per item and 512 values per token, the model only needs to store 2048 embeddings, 40\% less compared to the one-token-per-item mode \se{(see also Table~\ref{tb:memory_reduction})}.

\se{Figure~\ref{fig:multi-token} shows} that both metrics increase when the discretisation (number of possible values per token) increases. For example, with two tokens per item and 128 values per token, the model only reaches NDCG@10 of 0.0914. When the number of values per token increases to 2048, NDCG@10 increases to 0.124, \craig{thereby outperforming SASRec}. From the figure, we also see that four tokens per item are better in most cases than two tokens per item, except for the case with 2048 values per token -- in this case, the models demonstrate equal performance.  

Overall in answer to RQ2, we say that GPTRec in multi-token-per-item mode achieves comparable results with SASRec, but worse than GPTRec in one-token-per-item mode. However, in muti-token-per-item mode, it consumes less GPU memory, which \scr{is} important when working with larger datasets. 

\subsection{RQ3. Effect of cutoff K in Next-K generation}
\begin{figure*}
    \subfloat[Absolute]{
        \label{fig:next_k_vs_top_k_absolute}
        \includegraphics[width=0.4\textwidth]{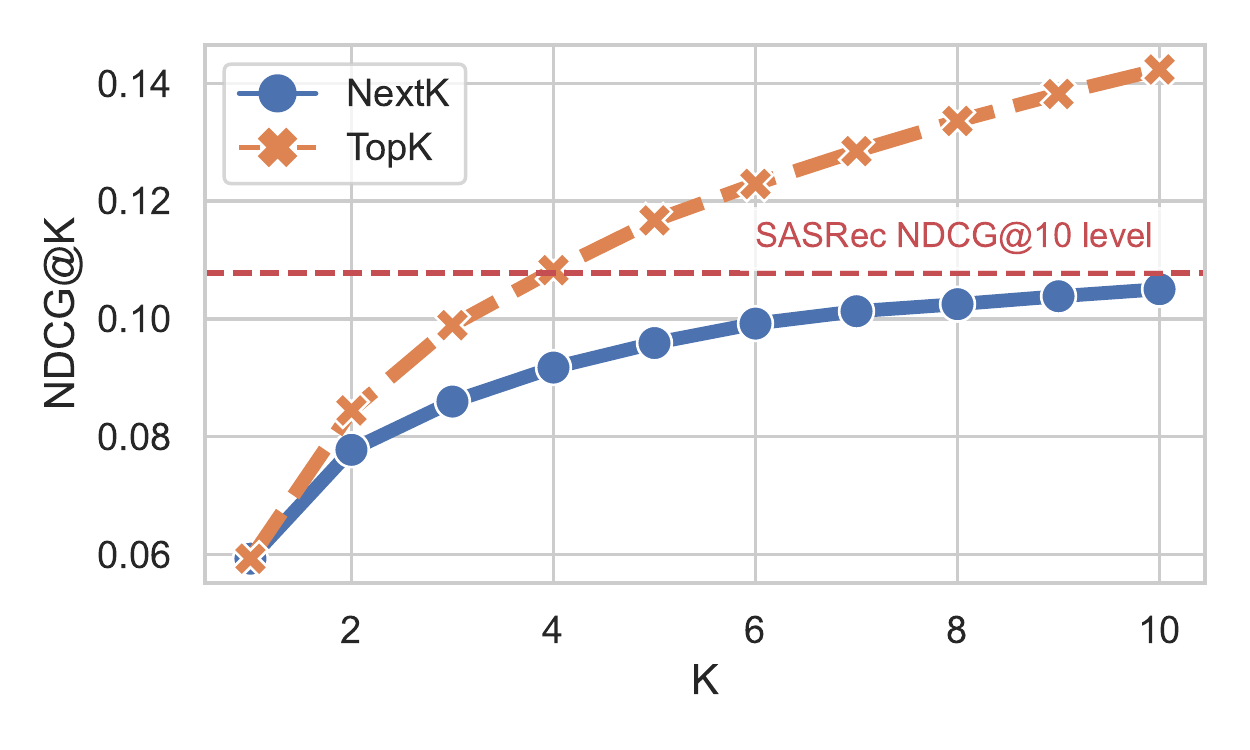}\vspace{-\baselineskip}}{
    }\subfloat[Relative]{
        \label{fig:next_k_vs_top_k_relative}
        \includegraphics[width=0.4\textwidth]{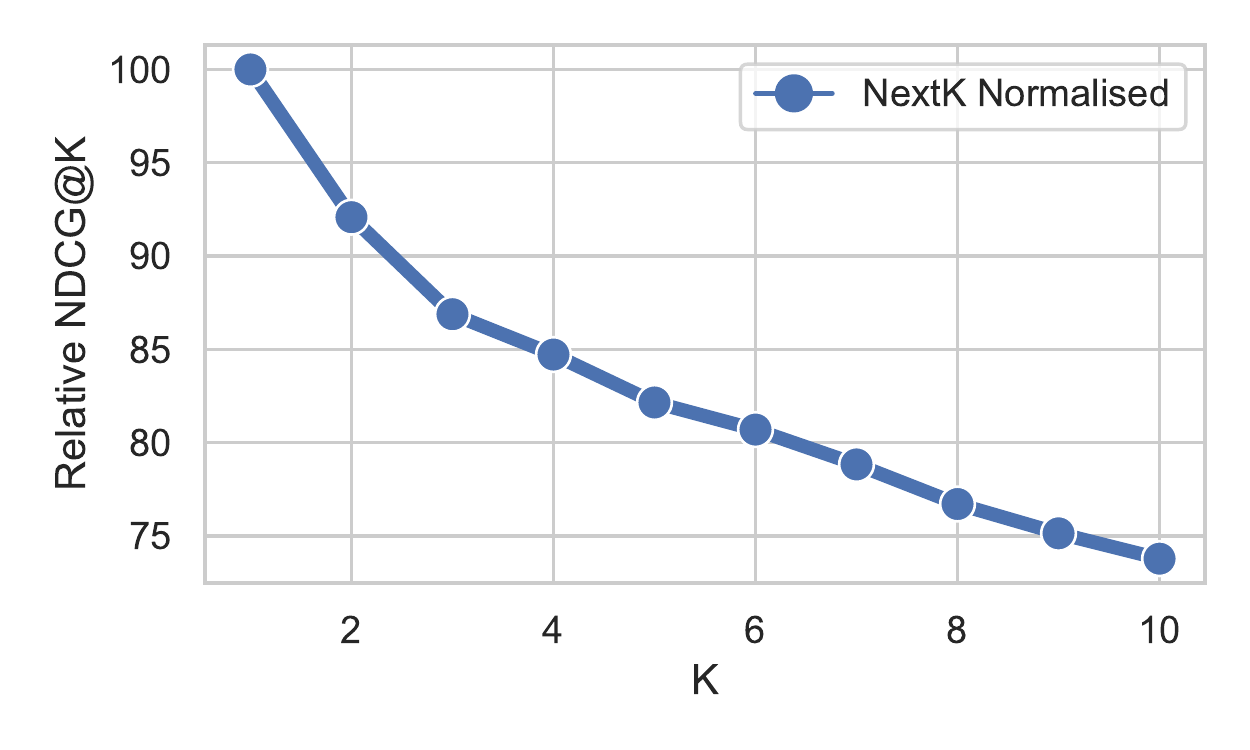}\vspace{-\baselineskip}}{
    }
    \caption{Comparison of GPTRec quality with Top-K and Next-K recommendation generation strategies. (a) shows the absolute value of the NDCG@K metric at different cutoff K.
    (b) shows the normalised quality of the Next-K strategy as the percentage of the Top-K strategy at the same cutoff K. } \label{fig:top_k_vs_next_k} 

\end{figure*}

We \scr{now} compare GPTRec performance with Top-K and Next-K recommendation strategies in a one-token-per-item mode to answer our last research question. \scr{Figure~\ref{fig:top_k_vs_next_k} compares these two-generation strategies: Figure~\ref{fig:next_k_vs_top_k_absolute} shows the absolute value of the NDCG@K metric at different cutoffs, whereas Figure~\ref{fig:next_k_vs_top_k_relative} shows the relative value of  NDCG@K metric of the Next-K strategy, using Top-K as a baseline. On analysis of Figure~\ref{fig:next_k_vs_top_k_relative}}, we observe that at cutoff $K=1$, \scr{GPTRec with Top-K strategy and GPTRec with Next-K strategy have the same performance}. Indeed, when $K=1$, both strategies generate the item with the highest probability and therefore are equal. 
However, with higher cutoffs, Next-K performs worse, decreasing gradually decreasing to 75\% of the Top-K quality at cutoff K=10; \scr{this is} an expected effect as the model in our experiment was not specifically tuned for Next-K recommendation, and therefore its training task is not aligned with the generation task. \scr{Nevertheless, as we see from Figure~\ref{fig:next_k_vs_top_k_absolute}}, even at cutoff K=10, \scr{GPTRec with Next-K strategy} performs similarly to the SASRec model (0.108 NDCG@10 in both cases), which means that this model can be a strong starting point for further tuning using  reinforcement learning or other techniques.

\craig{In summary, for RQ3, we find that} at low cutoffs, Next-K and Top-K strategies \scr{used with GPTRec} demonstrate similar performance. However, with increased K, the quality of the Next-K strategy gradually degrades. \scr{Nevertheless, GPTRec with the Next-K strategy}  retains competitive quality even with K=10, forming a strong basis for future tuning.

\section{Conclusions and Future Work} \label{sec:conclusion}
\balance
This paper presented GPTRec, a generative transformer model for the sequential recommendation problem \sdm{that supports a novel GPU memory-efficient SVD tokenisation and a novel Next-K recommendation generation strategy suitable for complex interdependent objectives. }
 This paper presented only early experiments with GPTRec and generative recommendations in general. In future work, we plan to expand model evaluation to more datasets (specifically large datasets, where multi-token-per-item promises the largest gains in memory). \se{Additionally, we plan to explore more advanced generation techniques, such as Beam Search, to mitigate the diversity problems in the multi-token-per-item generation.} We also plan to further research techniques to tune models for the Next-K generation and tune the model for complex objectives, such as diversity and  coverage. In addition, we plan to evaluate the model using more hyperparameter settings. 
\se{Another line of future work includes adaptations of proposed methods to the field of Generative IR. For example, adaptations of the proposed SVD tokenisation method can help resolve large document id vocabulary issues in Generative IR tasks. Similarly, \scr{GPTRec with} the Next-K strategy may help resolve complex search results diversity and fairness problems.}

\craig{To summarise this paper's findings}, we showed that in the simplest case, when the model is used in  one-token-per-item mode and uses a Top-K recommendation generation strategy, it \craig{showed} strong performance similar to BERT4Rec (4\% lower NDCG@10 on MovieLens-1M dataset).  Compared to existing models, such as BERT4Rec, the model supports a multi-token-per-item recommendation mode, which reduces memory footprint with large detests while retaining competitive quality (e.g.\ we showed that using an SVD-based tokeniser, the model can match SASRec's performance while storing  40\% fewer embeddings).  We also presented the Next-K recommendation generation strategy, a promising direction for tuning models for complex training objectives that include diversity and coverage. We showed that GPTRec could, in principle, be used with Next-K recommendations. While we observed quality degradation compared with traditional Top-K recommendations, the recommendation quality remains competitive with SASRec. Overall we conclude that generative sequential recommendation is a promising research direction, which can be beneficial for reducing the model's memory consumption and be advantageous for complex training objectives. We also believe that the ideas presented in this paper can \sd{benefit research areas beyond sequential recommender systems}, such as \sd{Generative} IR.

\FloatBarrier
\bibliographystyle{ACM-Reference-Format}
\bibliography{references}


\begin{thebibliography}{37}


\ifx \showCODEN    \undefined \def \showCODEN     #1{\unskip}     \fi
\ifx \showDOI      \undefined \def \showDOI       #1{#1}\fi
\ifx \showISBNx    \undefined \def \showISBNx     #1{\unskip}     \fi
\ifx \showISBNxiii \undefined \def \showISBNxiii  #1{\unskip}     \fi
\ifx \showISSN     \undefined \def \showISSN      #1{\unskip}     \fi
\ifx \showLCCN     \undefined \def \showLCCN      #1{\unskip}     \fi
\ifx \shownote     \undefined \def \shownote      #1{#1}          \fi
\ifx \showarticletitle \undefined \def \showarticletitle #1{#1}   \fi
\ifx \showURL      \undefined \def \showURL       {\relax}        \fi
\providecommand\bibfield[2]{#2}
\providecommand\bibinfo[2]{#2}
\providecommand\natexlab[1]{#1}
\providecommand\showeprint[2][]{arXiv:#2}

\bibitem[Brown et~al\mbox{.}(2020)]%
        {brownLanguageModelsAre2020}
\bibfield{author}{\bibinfo{person}{Tom Brown}, \bibinfo{person}{Benjamin Mann},
  \bibinfo{person}{Nick Ryder}, \bibinfo{person}{Melanie Subbiah},
  \bibinfo{person}{Jared~D Kaplan}, \bibinfo{person}{Prafulla Dhariwal},
  \bibinfo{person}{Arvind Neelakantan}, \bibinfo{person}{Pranav Shyam},
  \bibinfo{person}{Girish Sastry}, \bibinfo{person}{Amanda Askell},
  \bibinfo{person}{Sandhini Agarwal}, \bibinfo{person}{Ariel {Herbert-Voss}},
  \bibinfo{person}{Gretchen Krueger}, \bibinfo{person}{Tom Henighan},
  \bibinfo{person}{Rewon Child}, \bibinfo{person}{Aditya Ramesh},
  \bibinfo{person}{Daniel Ziegler}, \bibinfo{person}{Jeffrey Wu},
  \bibinfo{person}{Clemens Winter}, \bibinfo{person}{Chris Hesse},
  \bibinfo{person}{Mark Chen}, \bibinfo{person}{Eric Sigler},
  \bibinfo{person}{Mateusz Litwin}, \bibinfo{person}{Scott Gray},
  \bibinfo{person}{Benjamin Chess}, \bibinfo{person}{Jack Clark},
  \bibinfo{person}{Christopher Berner}, \bibinfo{person}{Sam McCandlish},
  \bibinfo{person}{Alec Radford}, \bibinfo{person}{Ilya Sutskever}, {and}
  \bibinfo{person}{Dario Amodei}.} \bibinfo{year}{2020}\natexlab{}.
\newblock \showarticletitle{Language {{Models}} Are {{Few-Shot Learners}}}. In
  \bibinfo{booktitle}{\emph{Proc. {{NeurIPS}}}}, Vol.~\bibinfo{volume}{33}.
  \bibinfo{pages}{1877--1901}.
\newblock


\bibitem[Cai et~al\mbox{.}(2022)]%
        {caiAspectRedistributionLearning2022}
\bibfield{author}{\bibinfo{person}{Wei Cai}, \bibinfo{person}{Weike Pan},
  \bibinfo{person}{Jingwen Mao}, \bibinfo{person}{Zhechao Yu}, {and}
  \bibinfo{person}{Congfu Xu}.} \bibinfo{year}{2022}\natexlab{}.
\newblock \showarticletitle{Aspect {{Re-distribution}} for {{Learning Better
  Item Embeddings}} in {{Sequential Recommendation}}}. In
  \bibinfo{booktitle}{\emph{Proc. {{RecSys}}}}. \bibinfo{pages}{49--58}.
\newblock


\bibitem[Carbonell and Goldstein(1998)]%
        {carbonellUseMMRDiversitybased1998}
\bibfield{author}{\bibinfo{person}{Jaime Carbonell} {and} \bibinfo{person}{Jade
  Goldstein}.} \bibinfo{year}{1998}\natexlab{}.
\newblock \showarticletitle{The Use of {{MMR}}, Diversity-Based Reranking for
  Reordering Documents and Producing Summaries}. In
  \bibinfo{booktitle}{\emph{Proc. {{SIGIR}}}}. \bibinfo{publisher}{{ACM}},
  \bibinfo{address}{{Melbourne Australia}}, \bibinfo{pages}{335--336}.
\newblock


\bibitem[Cho et~al\mbox{.}(2014)]%
        {choPropertiesNeuralMachine2014}
\bibfield{author}{\bibinfo{person}{Kyunghyun Cho}, \bibinfo{person}{Bart {van
  Merrienboer}}, \bibinfo{person}{Dzmitry Bahdanau}, {and}
  \bibinfo{person}{Yoshua Bengio}.} \bibinfo{year}{2014}\natexlab{}.
\newblock \bibinfo{title}{On the {{Properties}} of {{Neural Machine
  Translation}}: {{Encoder-Decoder Approaches}}}.
\newblock
\newblock
\showeprint[arxiv]{arXiv:1409.1259}


\bibitem[Cui et~al\mbox{.}(2022)]%
        {cuiM6RecGenerativePretrained2022}
\bibfield{author}{\bibinfo{person}{Zeyu Cui}, \bibinfo{person}{Jianxin Ma},
  \bibinfo{person}{Chang Zhou}, \bibinfo{person}{Jingren Zhou}, {and}
  \bibinfo{person}{Hongxia Yang}.} \bibinfo{year}{2022}\natexlab{}.
\newblock \bibinfo{title}{M6-{{Rec}}: {{Generative Pretrained Language Models}}
  Are {{Open-Ended Recommender Systems}}}.
\newblock
\newblock
\showeprint[arxiv]{2205.08084}~[cs]


\bibitem[Devlin et~al\mbox{.}(2019)]%
        {BERT}
\bibfield{author}{\bibinfo{person}{Jacob Devlin}, \bibinfo{person}{Ming-Wei
  Chang}, \bibinfo{person}{Kenton Lee}, {and} \bibinfo{person}{Kristina
  Toutanova}.} \bibinfo{year}{2019}\natexlab{}.
\newblock \showarticletitle{{{BERT}}: {{Pre-training}} of {{Deep Bidirectional
  Transformers}} for {{Language Understanding}}}. In
  \bibinfo{booktitle}{\emph{Proc. of {{NAACL-HLT}}}}.
  \bibinfo{pages}{4171--4186}.
\newblock


\bibitem[Du et~al\mbox{.}(2022)]%
        {CBiT}
\bibfield{author}{\bibinfo{person}{Hanwen Du}, \bibinfo{person}{Hui Shi},
  \bibinfo{person}{Pengpeng Zhao}, \bibinfo{person}{Deqing Wang},
  \bibinfo{person}{Victor~S. Sheng}, \bibinfo{person}{Yanchi Liu},
  \bibinfo{person}{Guanfeng Liu}, {and} \bibinfo{person}{Lei Zhao}.}
  \bibinfo{year}{2022}\natexlab{}.
\newblock \showarticletitle{Contrastive {{Learning}} with {{Bidirectional
  Transformers}} for {{Sequential Recommendation}}}. In
  \bibinfo{booktitle}{\emph{Proc. {{CIKM}}}}. \bibinfo{pages}{396--405}.
\newblock


\bibitem[Geng et~al\mbox{.}(2022)]%
        {gengRecommendationLanguageProcessing2022}
\bibfield{author}{\bibinfo{person}{Shijie Geng}, \bibinfo{person}{Shuchang
  Liu}, \bibinfo{person}{Zuohui Fu}, \bibinfo{person}{Yingqiang Ge}, {and}
  \bibinfo{person}{Yongfeng Zhang}.} \bibinfo{year}{2022}\natexlab{}.
\newblock \showarticletitle{Recommendation as {{Language Processing}}
  ({{RLP}}): {{A Unified Pretrain}}, {{Personalized Prompt}} \& {{Predict
  Paradigm}} ({{P5}})}. In \bibinfo{booktitle}{\emph{Proc. {{RecSys}}}}.
  \bibinfo{pages}{299--315}.
\newblock


\bibitem[Goodfellow et~al\mbox{.}(2016)]%
        {goodfellowDeepLearning2016}
\bibfield{author}{\bibinfo{person}{Ian Goodfellow}, \bibinfo{person}{Yoshua
  Bengio}, {and} \bibinfo{person}{Aaron Courville}.}
  \bibinfo{year}{2016}\natexlab{}.
\newblock \bibinfo{booktitle}{\emph{Deep {{Learning}}}}.
\newblock \bibinfo{publisher}{{MIT Press}}.
\newblock


\bibitem[Harper and Konstan(2015)]%
        {harperMovieLensDatasetsHistory2015}
\bibfield{author}{\bibinfo{person}{F.~Maxwell Harper} {and}
  \bibinfo{person}{Joseph~A. Konstan}.} \bibinfo{year}{2015}\natexlab{}.
\newblock \showarticletitle{The {{MovieLens Datasets}}: {{History}} and
  {{Context}}}.
\newblock \bibinfo{journal}{\emph{ACM Transactions on Interactive Intelligent
  Systems (TiiS)}} \bibinfo{volume}{5}, \bibinfo{number}{4}
  (\bibinfo{date}{Dec.} \bibinfo{year}{2015}), \bibinfo{pages}{19:1--19:19}.
\newblock
\showISSN{2160-6455}


\bibitem[Hidasi et~al\mbox{.}(2016)]%
        {GRU4Rec}
\bibfield{author}{\bibinfo{person}{Bal{\'a}zs Hidasi},
  \bibinfo{person}{Alexandros Karatzoglou}, \bibinfo{person}{Linas Baltrunas},
  {and} \bibinfo{person}{Domonkos Tikk}.} \bibinfo{year}{2016}\natexlab{}.
\newblock \showarticletitle{Session-Based {{Recommendations}} with {{Recurrent
  Neural Networks}}}. In \bibinfo{booktitle}{\emph{Proc. {{ICLR}}}}.
\newblock


\bibitem[Jaenich et~al\mbox{.}(2023)]%
        {jaenichColBERTFairPRFFairPseudoRelevance2023}
\bibfield{author}{\bibinfo{person}{Thomas Jaenich}, \bibinfo{person}{Graham
  McDonald}, {and} \bibinfo{person}{Iadh Ounis}.}
  \bibinfo{year}{2023}\natexlab{}.
\newblock \showarticletitle{{{ColBERT-FairPRF}}: {{Towards Fair
  Pseudo-Relevance Feedback}} in~{{Dense Retrieval}}}. In
  \bibinfo{booktitle}{\emph{Proc. {{ECIR}}}}. \bibinfo{pages}{457--465}.
\newblock


\bibitem[Kang and McAuley(2018)]%
        {SASRec}
\bibfield{author}{\bibinfo{person}{Wang-Cheng Kang} {and}
  \bibinfo{person}{Julian McAuley}.} \bibinfo{year}{2018}\natexlab{}.
\newblock \showarticletitle{Self-{{Attentive Sequential Recommendation}}}. In
  \bibinfo{booktitle}{\emph{Proc. {{ICDM}}}}. \bibinfo{pages}{197--206}.
\newblock
\showISSN{2374-8486}


\bibitem[Lee et~al\mbox{.}(2021)]%
        {leeDifferentiableRankingMetric2021}
\bibfield{author}{\bibinfo{person}{Hyunsung Lee}, \bibinfo{person}{Sangwoo
  Cho}, \bibinfo{person}{Yeongjae Jang}, \bibinfo{person}{Jaekwang Kim}, {and}
  \bibinfo{person}{Honguk Woo}.} \bibinfo{year}{2021}\natexlab{}.
\newblock \showarticletitle{Differentiable {{Ranking Metric Using Relaxed
  Sorting}} for {{Top-K Recommendation}}}.
\newblock \bibinfo{journal}{\emph{Proc. IEEE Access}}  \bibinfo{volume}{9}
  (\bibinfo{year}{2021}), \bibinfo{pages}{114649--114658}.
\newblock


\bibitem[Mehta et~al\mbox{.}(2022)]%
        {mehtaDSIUpdatingTransformer2022}
\bibfield{author}{\bibinfo{person}{Sanket~Vaibhav Mehta}, \bibinfo{person}{Jai
  Gupta}, \bibinfo{person}{Yi Tay}, \bibinfo{person}{Mostafa Dehghani},
  \bibinfo{person}{Vinh~Q. Tran}, \bibinfo{person}{Jinfeng Rao},
  \bibinfo{person}{Marc Najork}, \bibinfo{person}{Emma Strubell}, {and}
  \bibinfo{person}{Donald Metzler}.} \bibinfo{year}{2022}\natexlab{}.
\newblock \bibinfo{title}{{{DSI}}++: {{Updating Transformer Memory}} with {{New
  Documents}}}.
\newblock
\newblock
\showeprint[arxiv]{2212.09744}~[cs]


\bibitem[Ouyang et~al\mbox{.}(2022)]%
        {ouyangTrainingLanguageModels2022}
\bibfield{author}{\bibinfo{person}{Long Ouyang}, \bibinfo{person}{Jeff Wu},
  \bibinfo{person}{Xu Jiang}, \bibinfo{person}{Diogo Almeida},
  \bibinfo{person}{Carroll~L. Wainwright}, \bibinfo{person}{Pamela Mishkin},
  \bibinfo{person}{Chong Zhang}, \bibinfo{person}{Sandhini Agarwal},
  \bibinfo{person}{Katarina Slama}, \bibinfo{person}{Alex Ray},
  \bibinfo{person}{John Schulman}, \bibinfo{person}{Jacob Hilton},
  \bibinfo{person}{Fraser Kelton}, \bibinfo{person}{Luke Miller},
  \bibinfo{person}{Maddie Simens}, \bibinfo{person}{Amanda Askell},
  \bibinfo{person}{Peter Welinder}, \bibinfo{person}{Paul Christiano},
  \bibinfo{person}{Jan Leike}, {and} \bibinfo{person}{Ryan Lowe}.}
  \bibinfo{year}{2022}\natexlab{}.
\newblock \bibinfo{title}{Training Language Models to Follow Instructions with
  Human Feedback}.
\newblock
\newblock
\showeprint[arxiv]{2203.02155}~[cs]


\bibitem[Petrov and Macdonald(2022a)]%
        {PetrovRSS22}
\bibfield{author}{\bibinfo{person}{Aleksandr Petrov} {and}
  \bibinfo{person}{Craig Macdonald}.} \bibinfo{year}{2022}\natexlab{a}.
\newblock \showarticletitle{Effective and {{Efficient Training}} for
  {{Sequential Recommendation}} Using {{Recency Sampling}}}. In
  \bibinfo{booktitle}{\emph{Proc. {{RecSys}}}}. \bibinfo{pages}{81--91}.
\newblock


\bibitem[Petrov and Macdonald(2022b)]%
        {Bert4RecRepro}
\bibfield{author}{\bibinfo{person}{Aleksandr Petrov} {and}
  \bibinfo{person}{Craig Macdonald}.} \bibinfo{year}{2022}\natexlab{b}.
\newblock \showarticletitle{A {{Systematic Review}} and {{Replicability Study}}
  of {{BERT4Rec}} for {{Sequential Recommendation}}}. In
  \bibinfo{booktitle}{\emph{Proc. {{RecSys}}}}. \bibinfo{pages}{436--447}.
\newblock


\bibitem[Qiu et~al\mbox{.}(2022)]%
        {DuoRec}
\bibfield{author}{\bibinfo{person}{Ruihong Qiu}, \bibinfo{person}{Zi Huang},
  \bibinfo{person}{Hongzhi Yin}, {and} \bibinfo{person}{Zijian Wang}.}
  \bibinfo{year}{2022}\natexlab{}.
\newblock \showarticletitle{Contrastive {{Learning}} for {{Representation
  Degeneration Problem}} in {{Sequential Recommendation}}}. In
  \bibinfo{booktitle}{\emph{Proc. {{WSDM}}}}. \bibinfo{pages}{813--823}.
\newblock


\bibitem[Radford et~al\mbox{.}({[n.\,d.]})]%
        {radfordImprovingLanguageUnderstanding}
\bibfield{author}{\bibinfo{person}{Alec Radford}, \bibinfo{person}{Karthik
  Narasimhan}, \bibinfo{person}{Tim Salimans}, {and} \bibinfo{person}{Ilya
  Sutskever}.} \bibinfo{year}{[n.\,d.]}\natexlab{}.
\newblock \showarticletitle{Improving {{Language Understanding}} by
  {{Generative Pre-Training}}}.
\newblock  (\bibinfo{year}{[n.\,d.]}).
\newblock


\bibitem[Radford et~al\mbox{.}(2019)]%
        {gpt2}
\bibfield{author}{\bibinfo{person}{Alec Radford}, \bibinfo{person}{Jeffrey Wu},
  \bibinfo{person}{Rewon Child}, \bibinfo{person}{David Luan},
  \bibinfo{person}{Dario Amodei}, {and} \bibinfo{person}{Ilya Sutskever}.}
  \bibinfo{year}{2019}\natexlab{}.
\newblock \showarticletitle{Language {{Models}} Are {{Unsupervised Multitask
  Learners}}}.
\newblock \bibinfo{journal}{\emph{OpenAI blog}} (\bibinfo{year}{2019}).
\newblock


\bibitem[Raffel et~al\mbox{.}(2020)]%
        {raffelExploringLimitsTransfer2020}
\bibfield{author}{\bibinfo{person}{Colin Raffel}, \bibinfo{person}{Noam
  Shazeer}, \bibinfo{person}{Adam Roberts}, \bibinfo{person}{Katherine Lee},
  \bibinfo{person}{Sharan Narang}, \bibinfo{person}{Michael Matena},
  \bibinfo{person}{Yanqi Zhou}, \bibinfo{person}{Wei Li}, {and}
  \bibinfo{person}{Peter~J. Liu}.} \bibinfo{year}{2020}\natexlab{}.
\newblock \showarticletitle{Exploring the {{Limits}} of {{Transfer Learning}}
  with a {{Unified Text-to-Text Transformer}}}.
\newblock \bibinfo{journal}{\emph{Journal of Machine Learning Research}}
  \bibinfo{volume}{21}, \bibinfo{number}{140} (\bibinfo{year}{2020}),
  \bibinfo{pages}{1--67}.
\newblock
\showISSN{1533-7928}


\bibitem[{Rajput, Shashank} et~al\mbox{.}(2023)]%
        {rajputshashankRecommenderSystemsGenerative}
\bibfield{author}{\bibinfo{person}{{Rajput, Shashank}},
  \bibinfo{person}{{Mehta, Nikhil}}, \bibinfo{person}{{Singh, Anima}},
  \bibinfo{person}{{Keshavan, Raghunandan}}, \bibinfo{person}{{Vu, Trung}},
  \bibinfo{person}{{Heldt, Lukasz}}, \bibinfo{person}{{Hong, Lichan}},
  \bibinfo{person}{{Tay, Yi}}, \bibinfo{person}{{Tran, Vinh Q.}},
  \bibinfo{person}{{Samost, Jonah}}, \bibinfo{person}{{Kula, Maciej}},
  \bibinfo{person}{{Chi, Ed H.}}, {and} \bibinfo{person}{{Sathiamoorthy,
  Maheswaran}}.} \bibinfo{year}{2023}\natexlab{}.
\newblock \bibinfo{title}{Recommender {{Systems}} with {{Generative
  Retrieval}}}.
\newblock
\newblock


\bibitem[Sener and Koltun(2018)]%
        {sener2018multi}
\bibfield{author}{\bibinfo{person}{Ozan Sener} {and} \bibinfo{person}{Vladlen
  Koltun}.} \bibinfo{year}{2018}\natexlab{}.
\newblock \showarticletitle{Multi-task learning as multi-objective
  optimization}. In \bibinfo{booktitle}{\emph{Proc. {{NeurIPS}}}}.
\newblock


\bibitem[Sennrich et~al\mbox{.}(2016)]%
        {sennrichNeuralMachineTranslation2016}
\bibfield{author}{\bibinfo{person}{Rico Sennrich}, \bibinfo{person}{Barry
  Haddow}, {and} \bibinfo{person}{Alexandra Birch}.}
  \bibinfo{year}{2016}\natexlab{}.
\newblock \bibinfo{title}{Neural {{Machine Translation}} of {{Rare Words}} with
  {{Subword Units}}}.
\newblock
\newblock
\showeprint[arxiv]{1508.07909}~[cs]


\bibitem[Sun et~al\mbox{.}(2019)]%
        {BERT4Rec}
\bibfield{author}{\bibinfo{person}{Fei Sun}, \bibinfo{person}{Jun Liu},
  \bibinfo{person}{Jian Wu}, \bibinfo{person}{Changhua Pei},
  \bibinfo{person}{Xiao Lin}, \bibinfo{person}{Wenwu Ou}, {and}
  \bibinfo{person}{Peng Jiang}.} \bibinfo{year}{2019}\natexlab{}.
\newblock \showarticletitle{{{BERT4Rec}}: {{Sequential Recommendation}} with
  {{Bidirectional Encoder Representations}} from {{Transformer}}}. In
  \bibinfo{booktitle}{\emph{Proc. {{CIKM}}}}. \bibinfo{pages}{1441--1450}.
\newblock


\bibitem[Sun et~al\mbox{.}(2023)]%
        {sunLearningTokenizeGenerative2023}
\bibfield{author}{\bibinfo{person}{Weiwei Sun}, \bibinfo{person}{Lingyong Yan},
  \bibinfo{person}{Zheng Chen}, \bibinfo{person}{Shuaiqiang Wang},
  \bibinfo{person}{Haichao Zhu}, \bibinfo{person}{Pengjie Ren},
  \bibinfo{person}{Zhumin Chen}, \bibinfo{person}{Dawei Yin},
  \bibinfo{person}{Maarten {de Rijke}}, {and} \bibinfo{person}{Zhaochun Ren}.}
  \bibinfo{year}{2023}\natexlab{}.
\newblock \bibinfo{title}{Learning to {{Tokenize}} for {{Generative
  Retrieval}}}.
\newblock
\newblock
\showeprint[arxiv]{2304.04171}~[cs]


\bibitem[Tang and Wang(2018)]%
        {Caser}
\bibfield{author}{\bibinfo{person}{Jiaxi Tang} {and} \bibinfo{person}{Ke
  Wang}.} \bibinfo{year}{2018}\natexlab{}.
\newblock \showarticletitle{Personalized {{Top-N Sequential Recommendation}}
  via {{Convolutional Sequence Embedding}}}. In \bibinfo{booktitle}{\emph{Proc.
  {{WSDM}}}}. \bibinfo{pages}{565--573}.
\newblock


\bibitem[Tay et~al\mbox{.}(2022)]%
        {tayTransformerMemoryDifferentiable}
\bibfield{author}{\bibinfo{person}{Yi Tay}, \bibinfo{person}{Vinh~Q. Tran},
  \bibinfo{person}{Mostafa Dehghani}, \bibinfo{person}{Jianmo Ni},
  \bibinfo{person}{Dara Bahri}, \bibinfo{person}{Harsh Mehta},
  \bibinfo{person}{Zhen Qin}, \bibinfo{person}{Kai Hui}, \bibinfo{person}{Zhe
  Zhao}, \bibinfo{person}{Jai Gupta}, \bibinfo{person}{Tal Schuster},
  \bibinfo{person}{William~W. Cohen}, {and} \bibinfo{person}{Donald Metzler}.}
  \bibinfo{year}{2022}\natexlab{}.
\newblock \showarticletitle{Transformer Memory as a Differentiable Search
  Index}.
\newblock  (\bibinfo{year}{2022}).
\newblock
\showeprint[arxiv]{2202.06991}~[cs.CL]


\bibitem[Vaswani et~al\mbox{.}(2017)]%
        {Transformer}
\bibfield{author}{\bibinfo{person}{Ashish Vaswani}, \bibinfo{person}{Noam
  Shazeer}, \bibinfo{person}{Niki Parmar}, \bibinfo{person}{Jakob Uszkoreit},
  \bibinfo{person}{Llion Jones}, \bibinfo{person}{Aidan~N Gomez},
  \bibinfo{person}{{\L}ukasz Kaiser}, {and} \bibinfo{person}{Illia
  Polosukhin}.} \bibinfo{year}{2017}\natexlab{}.
\newblock \showarticletitle{Attention Is {{All}} You {{Need}}}. In
  \bibinfo{booktitle}{\emph{Proc. {{NeurIPS}}}}.
\newblock


\bibitem[Wolf et~al\mbox{.}(2020)]%
        {wolfHuggingFaceTransformersStateoftheart2020}
\bibfield{author}{\bibinfo{person}{Thomas Wolf}, \bibinfo{person}{Lysandre
  Debut}, \bibinfo{person}{Victor Sanh}, \bibinfo{person}{Julien Chaumond},
  \bibinfo{person}{Clement Delangue}, \bibinfo{person}{Anthony Moi},
  \bibinfo{person}{Pierric Cistac}, \bibinfo{person}{Tim Rault},
  \bibinfo{person}{R{\'e}mi Louf}, \bibinfo{person}{Morgan Funtowicz},
  \bibinfo{person}{Joe Davison}, \bibinfo{person}{Sam Shleifer},
  \bibinfo{person}{Patrick {von Platen}}, \bibinfo{person}{Clara Ma},
  \bibinfo{person}{Yacine Jernite}, \bibinfo{person}{Julien Plu},
  \bibinfo{person}{Canwen Xu}, \bibinfo{person}{Teven~Le Scao},
  \bibinfo{person}{Sylvain Gugger}, \bibinfo{person}{Mariama Drame},
  \bibinfo{person}{Quentin Lhoest}, {and} \bibinfo{person}{Alexander~M. Rush}.}
  \bibinfo{year}{2020}\natexlab{}.
\newblock \bibinfo{title}{{{HuggingFace}}'s {{Transformers}}:
  {{State-of-the-art Natural Language Processing}}}.
\newblock
\newblock
\showeprint[arxiv]{1910.03771}~[cs]


\bibitem[Wu et~al\mbox{.}(2016)]%
        {wuGoogleNeuralMachine2016}
\bibfield{author}{\bibinfo{person}{Yonghui Wu}, \bibinfo{person}{Mike
  Schuster}, \bibinfo{person}{Zhifeng Chen}, \bibinfo{person}{Quoc~V. Le},
  \bibinfo{person}{Mohammad Norouzi}, \bibinfo{person}{Wolfgang Macherey},
  \bibinfo{person}{Maxim Krikun}, \bibinfo{person}{Yuan Cao},
  \bibinfo{person}{Qin Gao}, \bibinfo{person}{Klaus Macherey},
  \bibinfo{person}{Jeff Klingner}, \bibinfo{person}{Apurva Shah},
  \bibinfo{person}{Melvin Johnson}, \bibinfo{person}{Xiaobing Liu},
  \bibinfo{person}{{\L}ukasz Kaiser}, \bibinfo{person}{Stephan Gouws},
  \bibinfo{person}{Yoshikiyo Kato}, \bibinfo{person}{Taku Kudo},
  \bibinfo{person}{Hideto Kazawa}, \bibinfo{person}{Keith Stevens},
  \bibinfo{person}{George Kurian}, \bibinfo{person}{Nishant Patil},
  \bibinfo{person}{Wei Wang}, \bibinfo{person}{Cliff Young},
  \bibinfo{person}{Jason Smith}, \bibinfo{person}{Jason Riesa},
  \bibinfo{person}{Alex Rudnick}, \bibinfo{person}{Oriol Vinyals},
  \bibinfo{person}{Greg Corrado}, \bibinfo{person}{Macduff Hughes}, {and}
  \bibinfo{person}{Jeffrey Dean}.} \bibinfo{year}{2016}\natexlab{}.
\newblock \bibinfo{title}{Google's {{Neural Machine Translation System}}:
  {{Bridging}} the {{Gap}} between {{Human}} and {{Machine Translation}}}.
\newblock
\newblock
\showeprint[arxiv]{1609.08144}~[cs]


\bibitem[Xia et~al\mbox{.}(2015)]%
        {xiaLearningMaximalMarginal2015}
\bibfield{author}{\bibinfo{person}{Long Xia}, \bibinfo{person}{Jun Xu},
  \bibinfo{person}{Yanyan Lan}, \bibinfo{person}{Jiafeng Guo}, {and}
  \bibinfo{person}{Xueqi Cheng}.} \bibinfo{year}{2015}\natexlab{}.
\newblock \showarticletitle{Learning {{Maximal Marginal Relevance Model}} via
  {{Directly Optimizing Diversity Evaluation Measures}}}. In
  \bibinfo{booktitle}{\emph{Proc. {{SIGIR}}}}. \bibinfo{pages}{113--122}.
\newblock


\bibitem[Yuan et~al\mbox{.}(2019)]%
        {yuanSimpleConvolutionalGenerative2019}
\bibfield{author}{\bibinfo{person}{Fajie Yuan}, \bibinfo{person}{Alexandros
  Karatzoglou}, \bibinfo{person}{Ioannis Arapakis}, \bibinfo{person}{Joemon~M.
  Jose}, {and} \bibinfo{person}{Xiangnan He}.} \bibinfo{year}{2019}\natexlab{}.
\newblock \showarticletitle{A {{Simple Convolutional Generative Network}} for
  {{Next Item Recommendation}}}. In \bibinfo{booktitle}{\emph{Proc. {{WSDM}}}}.
  \bibinfo{pages}{582--590}.
\newblock


\bibitem[Zhou et~al\mbox{.}(2009)]%
        {zhou2009accurate}
\bibfield{author}{\bibinfo{person}{Tao Zhou}, \bibinfo{person}{Ri-Qi Su},
  \bibinfo{person}{Run-Ran Liu}, \bibinfo{person}{Luo-Luo Jiang},
  \bibinfo{person}{Bing-Hong Wang}, {and} \bibinfo{person}{Yi-Cheng Zhang}.}
  \bibinfo{year}{2009}\natexlab{}.
\newblock \showarticletitle{Accurate and diverse recommendations via
  eliminating redundant correlations}.
\newblock \bibinfo{journal}{\emph{New Journal of Physics}}
  \bibinfo{volume}{11}, \bibinfo{number}{12} (\bibinfo{year}{2009}).
\newblock


\bibitem[Zhou et~al\mbox{.}(2022)]%
        {zhouDynamicRetrieverPretrainedModelbased}
\bibfield{author}{\bibinfo{person}{Yu-Jia Zhou}, \bibinfo{person}{Jing Yao},
  \bibinfo{person}{Zhi-Cheng Dou}, \bibinfo{person}{Ledell Wu}, {and}
  \bibinfo{person}{Ji-Rong Wen}.} \bibinfo{year}{2022}\natexlab{}.
\newblock \showarticletitle{{{DynamicRetriever}}: {{A Pre-trained Model-based
  IR System Without}} an {{Explicit Index}}}.
\newblock \bibinfo{journal}{\emph{Machine Intelligence Research}}
  \bibinfo{volume}{20}, \bibinfo{number}{2} (\bibinfo{year}{2022}),
  \bibinfo{pages}{276--288}.
\newblock


\bibitem[Zhu et~al\mbox{.}(2019)]%
        {zhuImprovingTopKRecommendation2019}
\bibfield{author}{\bibinfo{person}{Ziwei Zhu}, \bibinfo{person}{Jianling Wang},
  {and} \bibinfo{person}{James Caverlee}.} \bibinfo{year}{2019}\natexlab{}.
\newblock \showarticletitle{Improving {{Top-K Recommendation}} via {{Joint
  Collaborative Autoencoders}}}. In \bibinfo{booktitle}{\emph{Proc. {{WWW}}}}.
\newblock


\end{thebibliography}
\end{document}